\documentclass[aps,twocolumn,showpacs]{revtex4}
\usepackage{amsmath}
\usepackage{epsf}
\usepackage{bm}
\begin{document}
\title{Exact Sequence Analysis for Three-Dimensional HP Lattice Proteins}
\author{Reinhard Schiemann}
\email[E-mail: ]{Reinhard.Schiemann@itp.uni-leipzig.de}
\author{Michael Bachmann}
\email[E-mail: ]{Michael.Bachmann@itp.uni-leipzig.de}
\author{Wolfhard Janke}
\email[E-mail: ]{Wolfhard.Janke@itp.uni-leipzig.de}
\homepage[\\ Homepage: ]{http://www.physik.uni-leipzig.de/CQT}
\affiliation{Institut f\"ur Theoretische Physik, Universit\"at Leipzig,
Augustusplatz 10/11, D-04109 Leipzig, Germany}
\begin{abstract}
We have exactly enumerated all sequences and conformations of HP proteins 
with chains of up to 19 monomers on the simple cubic lattice. For two variants
of the hydrophobic-polar (HP) model, where only two types of monomers
are distinguished, we determined and statistically analyzed designing 
sequences, i.e., sequences that have a non-degenerate ground state.
Furthermore we were interested in characteristic thermodynamic 
properties of HP proteins with designing sequences. In order 
to be able to perform these exact studies, we applied an efficient 
enumeration method based on contact sets.  
\end{abstract}
\maketitle
\section{Introduction}
\label{intro}
Real proteins are build up of sequences of amino acids covalently 
linked by peptide bonds. Twenty different types of amino acids
occurring in protein sequences are known. For a protein consisting of $N$ amino acid residues
there are thus in principle
$20^N$ possibilities to form sequences or primary protein structures. Single domain
polypeptides usually possess $N=30\ldots 400$ residues; proteins built
up of several domains can consist of up to 4000 amino acids. Only a few
of the $20^N$ possible proteins, however, are actually realized in nature and  
are functional in a sense that they fulfil a specific task in a biological
system. This requires the native structure of the protein
to be unique and stable against moderate fluctuations of the 
environmental chemical and physical conditions. It is widely believed
that the native state resides in a deep funnel-like minimum of
the free energy landscape~\cite{dill1}. Since the energy of a protein depends on
its sequence, it seems plausible that only such sequences of residues
are favored whose associated energy landscape shows up a pronounced 
global minimum. 
On the other hand, from the conformational point of view, it can be estimated 
that the number of structures proteins typically fold into, is only of order 
1000 -- this is at least two orders of magnitude less than the number of 
known proteins. 

Hence, exposing the nature of the relationship between sequences
(primary structure) and conformations (secondary and higher structures)
is one of the main aspects in protein research~\cite{tang1}. Attacking
this general problem by means of computer simulations based on realistic
interactions is currently impossible. There are two major reasons being
responsible for this. Firstly, the precise form of the energy function 
in an all-atom approach containing all molecular and nuclear interactions within
the polypeptide as well as the influence of the solvent is still under
consideration. An important question is what ``level of detail'' is necessary to 
model proteins in general. Considering an exemplified sequence of amino
acid residues, the use of different force fields usually leads to different 
predictions of the native state. Secondly, even if a reliable model
would exist, the sequence space is too large to be completely scanned 
by enumeration in order to search for the small number of sequences
with appropriate free energy landscape 
(the number of primary structures of very short 
peptides with, say only 10 residues, is $20^{10}\approx 10^{13}$).      

In order to have a chance to perform such an analysis, the model
must be drastically simplified. The simplest model to describe very
qualitatively the folding behavior of proteins is the HP model~\cite{dill2},
where the continuous conformational space is reduced to discrete regular lattices 
and conformations of proteins are modeled as self-avoiding walks restricted to 
the lattice. In this model
it is assumed that the hydrophobic interaction is the essential driving force towards a
native fold. It is expected that the hydrophobic side chains are screened
from the aqueous environment by hydrophilic residues. Therefore, the sequences of
HP proteins consist of only two types of monomers (or classes of 
amino acids), amino acids with high hydrophobicity are treated as 
hydrophobic monomers ($H$), while the class of polar (or hydrophilic) residues
is represented by polar monomers ($P$). In order to achieve the formation 
of a hydrophobic core surrounded by a shell of polar monomers, the interaction
between hydrophobic monomers is attractive in the standard formulation of the model.
All other interactions are neglected.
Variants of the HP model also take
into account
(weaker) interactions between $H$ and $P$ monomers as well as 
between polar monomers~\cite{tang1}. 
  
Although it is obvious that this model can describe the folding process very 
roughly only~\cite{shak1,shak2,vend1,chan1}, much work has
been done to find lowest-energy states and their degeneracy for given sequences, or
in the inverse problem, to identify all sequences of given length whose native 
conformation matches a given target structure. 
As simple as this model seems to be, it has been proven to be an NP-complete
problem in two and three dimensions~\cite{NPcompl}. Therefore, sophisticated
algorithms were applied to find lowest energy states for chains of up to 136 
monomers. The methods applied are based on very different algorithms, ranging 
from exact enumeration in two dimensions~\cite{irb1,irb2} and three dimensions
on cuboid (compact) lattices~\cite{tang1,tang2}, and hydrophobic core construction
methods~\cite{dill3,dill4} over genetic 
algorithms~\cite{unger1,kras1,cui1,lesh1,jiang1}, Monte Carlo simulations 
with different 
types of move sets~\cite{seno1,rama1,irb3,lee1}, and generalized ensemble 
approaches~\cite{iba1} to Rosenbluth
chain growth methods~\cite{rosen1} of the {\em 'Go with the Winners'} 
type~\cite{aldous1,grass1,grass2,hsu1,bj1,bj2}. With some of these algorithms, thermodynamic 
quantities of lattice heteropolymers were studied as well~\cite{iba1,grass2,bj1,bj2,naj1}. 
 
In this work, we apply an exact enumeration method to three-dimensional 
HP proteins being not necessarily compact on the simple cubic (s.c.) lattice.
For efficiency, we enumerated contact sets for chains of given length instead of conformations. 
In order to study the interplay between sequences and conformations and to investigate
peculiarities of designing sequences, we perform a statistical analysis of the complete 
spaces of conformations and sequences for chains of up to $N=19$ monomers. 

In Section~\ref{models} we give a review on the two variants of the HP model we
use in our study, the original HP model and a variant taking into account an additional
interaction between hydrophobic and polar residues. This is followed by Section~\ref{sawcm},
where we discuss self-avoiding conformations and contact sets. Then, in Section~\ref{seqana},
we perform an exact statistical analysis of properties of designing sequences and 
native conformations with lengths up to $19$ monomers in comparison with the bulk of 
all possibilities to form sequences and to generate conformations, respectively. Since   
the exact data obtained with our algorithm can be rearranged in terms of
the energy levels of the conformations, we are also able to determine the densities
of states for all sequences. This allows for the study of the energetic thermodynamic properties
of sequences whose associated ground state is unique or not, and their comparison from 
a thermodynamic point of view. We do just that in Section~\ref{sectdens}. The paper is
then concluded by summarising our results in Section~\ref{sum}.
\section{HP Models}
\label{models}
A monomer of a HP sequence $\bm{\sigma}=(\sigma_1,\sigma_2,\ldots,\sigma_N)$ 
is characterized by its residual type
($\sigma_i=P$ for polar and $\sigma_i=H$ for hydrophobic residues), 
the place $1\le i\le N$ within the chain of length $N$, and the spatial position ${\bf x}$
to be measured in units of the lattice spacing. A conformation is then symbolized by the
vector of the coordinates of successive monomers, 
${\bf X}=({\bf x}_1,{\bf x}_2,\ldots,{\bf x}_N)$.
We denote by $x_{ij}=|{\bf x}_i-{\bf x}_j|$
the distance between the $i$th and the $j$th monomer.
The bond length between adjacent monomers in the chain is identical with the spacing of the
used regular lattice with coordination number $k$. These covalent bonds are thus
not stretchable.
A monomer and its non-bonded nearest neighbors may form {\em contacts}.  
Therefore, the maximum number of contacts of a monomer within the chain is $(k-2)$ and
$(k-1)$ for the monomers at the ends of the chain. To account for the excluded volume,
lattice proteins are self-avoiding, i.e., two monomers cannot occupy the same 
lattice site. The general energy function of the non-covalent interactions reads 
in energy units $\varepsilon$ (we set $\varepsilon=1$ in the following)
\begin{equation} 
\label{hpgen} 
E=\varepsilon\sum\limits_{\langle i, j>i+1 \rangle} C_{ij}U_{\sigma_i\sigma_j},
\end{equation}
where $C_{ij}=(1-\delta_{i+1\,j})\Delta(x_{ij}-1)$ with 
\begin{equation}
\label{defDelta}
\Delta(z)=\left\{ \begin{array}{cl}
1, & \hspace{3mm} z=0,\\
0, & \hspace{3mm} z\neq 0
\end{array}\right. 
\end{equation}
is a symmetric $N\times N$ matrix called {\em contact map} and 
\begin{equation}
\label{intmatrix}
U_{\sigma_i\sigma_j}=\left(\begin{array}{cc}
u_{HH} & u_{HP}\\
u_{HP} & u_{PP} \end{array}\right)
\end{equation}
is the $2\times 2$ interaction matrix. Its elements $u_{\sigma_i\sigma_j}$ correspond to
the energy of $HH$, $HP$, and $PP$ contacts. For labeling purposes we shall adopt
the convention that $\sigma_i=0\,\hat{=}\, P$ and $\sigma_i=1\,\hat{=}\, H$.

In the simplest formulation~\cite{dill2}, which we will refer to as HP model in the 
following, only the attractive hydrophobic interaction is nonzero, $u^{\rm HP}_{HH}=-1$, while
$u^{\rm HP}_{HP}=u^{\rm HP}_{PP}=0$. 
Therefore, $U^{\rm HP}_{\sigma_i\sigma_j}=-\delta_{\sigma_i H}\delta_{\sigma_j H}$. This 
model has been extensively used to identify ground states of HP sequences, some of which are 
believed to show up qualitative properties comparable with realistic proteins whose 20-letter 
sequence was translated into the 2-letter code of the HP model~\cite{shak2,dill3,unger1,103lat,103toma}. 
As we will see later on, this simple form of the HP
model suffers, however, from the fact that the lowest-energy states are usually highly 
degenerate and therefore the number of designing sequences (i.e., sequences with unique 
ground state -- up to the usual translational, rotational, and reflection symmetries) 
is very small, at least on the simple cubic lattice.   

For a more reliable statistical sequence analysis, we compare with another model of HP type, as proposed in 
Ref.~\cite{tang1}. This model was motivated by results from an analysis of inter-residue contact energies 
between real amino acids~\cite{mj1}. To this end, an attractive nonzero energy contribution for
contacts between $H$ and $P$ monomers is assumed~\cite{tang1}. In what follows we call this the MHP ({\em mixed}
HP) model. The elements of the interaction matrix~(\ref{intmatrix}) are chosen to be
$u^{\rm MHP}_{HH}=-1$, $u^{\rm MHP}_{HP}=-1/2.3\approx -0.435$, and $u^{\rm MHP}_{PP}=0$ 
\cite{fnscale}. 
The additional
$H$-$P$ interaction breaks conformational symmetries yielding a much higher number of
designing sequences on cubic lattices.
\section{Self-Avoiding Walks and Contact Matrices}
\label{sawcm}
\begin{figure}
\centerline {
\epsfxsize =8.9cm \epsfbox {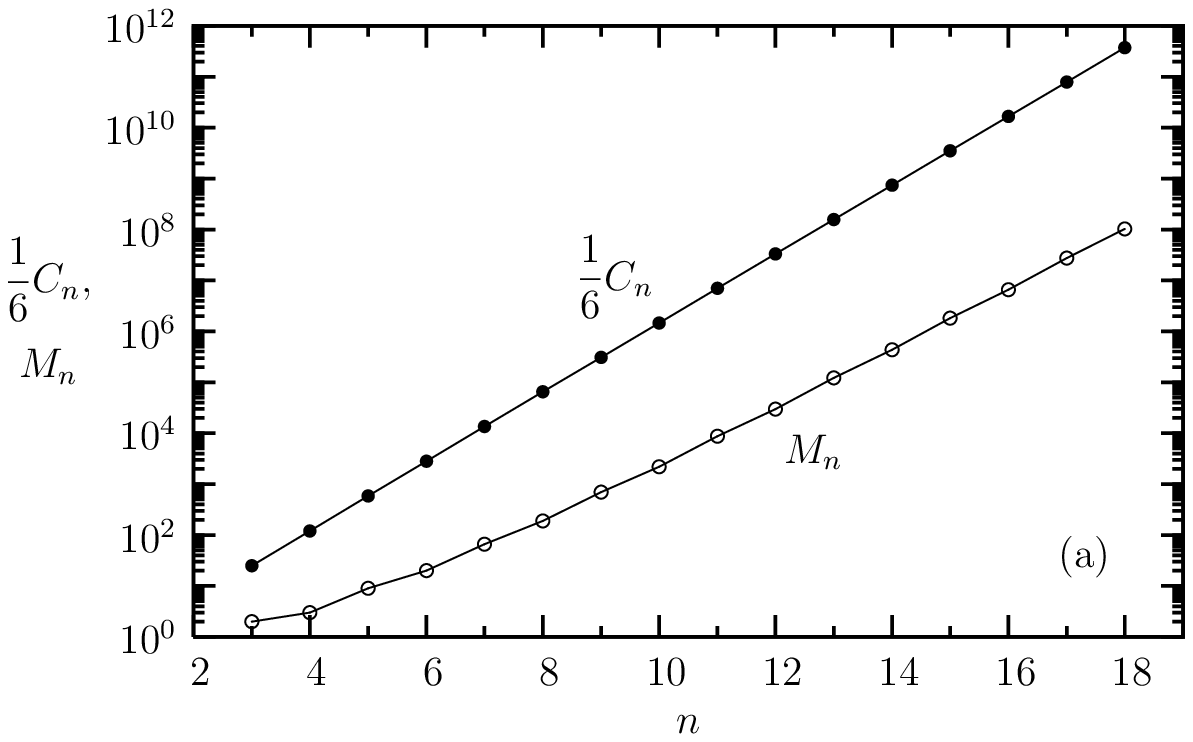}
}

\centerline {
\epsfxsize =8.5cm \epsfbox {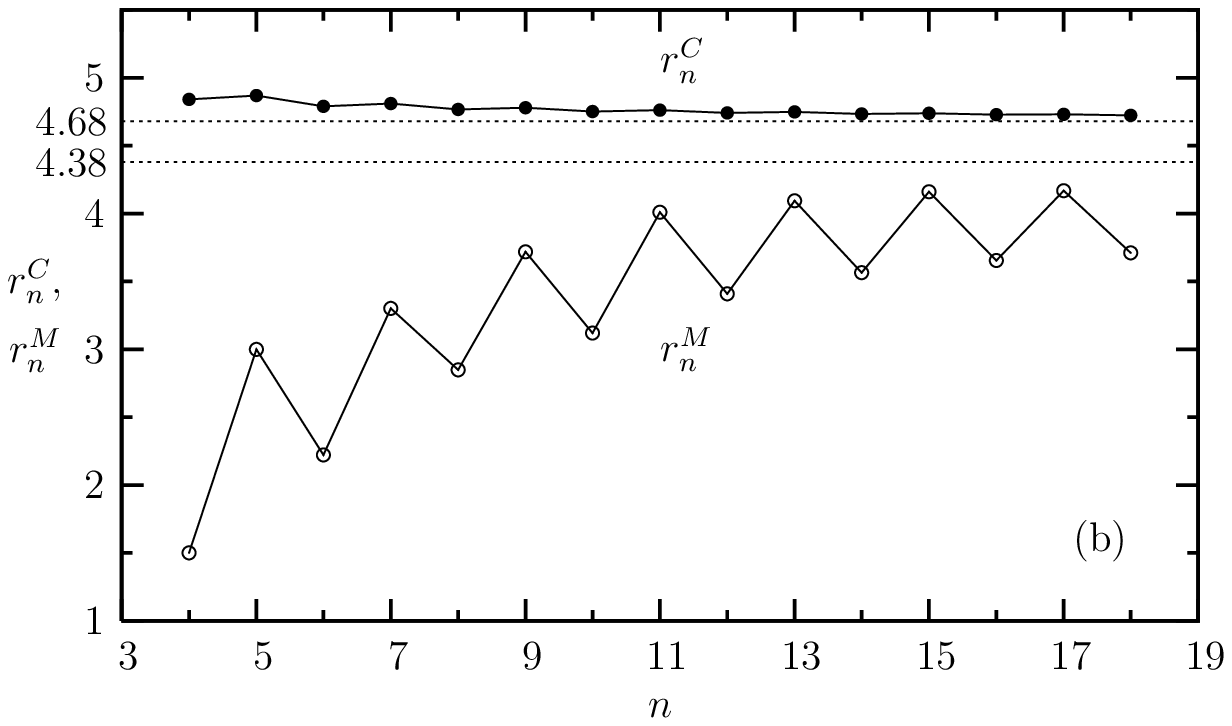}
}
\caption{\label{figconfs} (a) Dependence of the numbers of self-avoiding walks $C_n$ and contact matrices
$M_n$ on the number of steps $n=N-1$. (b) Ratios of numbers of self-avoiding walks $r^C_n=C_n/C_{n-1}$ and 
contact matrices $r^M_n=M_n/M_{n-1}$. The dotted lines indicate the values the
respective series converge to, $r^C_\infty=\mu_C\approx 4.68$ and $r^M_\infty=\mu_M\approx 4.38$,
respectively.
}
\end{figure}
Since lattice polymers are self-avoiding walks, the total number of conformations 
for a chain with $N$ monomers is not known exactly. For $N\to\infty$ it is widely
believed that in leading order the scaling law~\cite{gaunt1,gutt1}
\begin{equation}
\label{scalsaw}
C_n=A\mu_C^n n^{\gamma-1}
\end{equation}
holds, where $n$ is the number of self-avoiding steps (i.e., $n=N-1$).
In this expression, $\mu_C$ is the effective coordination number of the lattice, $\gamma$
is a universal exponent, and $A$ is a non-universal amplitude. In Table~\ref{tabconfs} we have listed
the exactly enumerated number for self-avoiding conformations of chains with up to $N=n+1=19$
monomers. Based on these data we estimated $\mu_C\approx 4.684$ and $\gamma\approx 1.16$ by extrapolating 
the results obtained with the ratio method~\cite{gaunt1,ratiom}. These results are in good agreement
with previous enumeration results~\cite{mac1,gutt2,chen1}, Monte Carlo methods~\cite{cara1} 
and field-theoretic estimates for $\gamma$~\cite{guida1}. 
We should note that it is not the aim of this work to extend the numbers of walks $C_N$ 
in Table~\ref{tabconfs} which has already been enumerated up to $n=26$ steps (and hence 
$C_{27}\approx 5.49\times 10^{17}$ self-avoiding conformations with $N=27$ monomers)~\cite{gutt2}. Rather, we
scan the combined space of HP sequences and conformations which contains
for chains of $N=19$ monomers $2^{19}C_{19}\approx 1.17\times 10^{18}$ possible combinations. Therefore,
the computational efforts in our study are comparably demanding. 
\begin{table}
\caption{\label{tabconfs} Number of conformations $C_N$ (with a global symmetry factor of 6 divided out)
and contact matrices $M_N$ for chains with $N$ monomers (or, equivalently, self-avoiding walks 
with $n=N-1$ steps).}
\centerline{
\begin{tabular}{rr@{\hspace{3mm}}r@{\hspace{3mm}}r@{\hspace{3mm}}r@{\hspace{5mm}}} \hline \hline
  $N$ & $n$ & \multicolumn{1}{c}{$\frac{1}{6}C_N$} & 
  \multicolumn{1}{c}{$M_N$} & \multicolumn{1}{c}{$\frac{1}{6}C_N / M_N$}\\
  \hline
  4 & 3 & 25 & 2 & 13\\
  5 & 4 & 121 & 3 & 40\\
  6 & 5 & 589 & 9 & 65\\
  7 & 6 & 2\,821 & 20 & 141\\
  8 & 7 & 13\,565 & 66 & 206\\
  9 & 8 & 64\,661 & 188 & 344\\
  10 & 9 & 308\,981 & 699 & 442\\
  11 & 10 & 1\,468\,313 & 2\,180 & 674\\
  12 & 11 & 6\,989\,025 & 8\,738 & 800\\
  13 & 12 & 33\,140\,457 & 29\,779 & 1\,113\\
  14 & 13 & 157\,329\,085 & 121\,872 & 1\,290\\
  15 & 14 & 744\,818\,613 & 434\,313 & 1\,715\\
  16 & 15 & 3\,529\,191\,009 & 1\,806\,495 & 1\,954\\
  17 & 16 & 16\,686\,979\,329 & 6\,601\,370 & 2\,528\\
  18 & 17 & 78\,955\,042\,017 & 27\,519\,000 & 2\,869\\
  19 & 18 & 372\,953\,947\,349 & 102\,111\,542 & 3\,652\\ \hline \hline
\end{tabular}}
\end{table}

In models with the general form~(\ref{hpgen}),
where the calculation of the energy reduces to the summation over contacts (i.e., pairs of monomers
being nearest neighbors on the lattice but nonadjacent along the chain) of a given conformation,
the number of conformations that must necessarily be enumerated can drastically be decreased by
considering only classes of conformations, so-called contact sets~\cite{irb2,domany1}. A contact set is uniquely 
characterized by a corresponding contact map (or contact matrix), but a single conformation is not.
Thus, for determining energetic quantities of different sequences, it is sufficient to carry out enumerations
over contact sets. In a first step, however, the contact sets and their degeneracy, i.e., the number
of conformations belonging to each set, must be determined and stored. Then, 
the loop over all non-redundant sequences is performed for all contact sets instead of conformations. The technical details of our implementation will be described elsewhere~\cite{sbj1}.

In Table~\ref{tabconfs}, the resulting
numbers of contact sets $M_N$ are summarized and, although also growing exponentially
(see Figs.~\ref{figconfs}(a) and (b)), the gain of efficiency by enumerating contact sets, is documented by the ratio 
between $C_N$ and $M_N$ in the last column. Assuming that 
the number of contact sets $M_n$ follows a scaling law
similar to Eq.~(\ref{scalsaw}), we estimated the effective coordination number to be
approximately $\mu_M\approx 4.38$. 
Unfortunately, the ratios of numbers of contact sets for even and odd numbers of walks
oscillate much stronger than for the number of conformations, as is shown in Fig.~\ref{figconfs}(b). 
This renders an accurate scaling analysis (in particular for the exponent $\gamma$)
based on the data for the relatively small number of steps much more difficult than
for self-avoiding walks. 
\section{Exact Statistical Analysis of Designing Sequences}
\label{seqana}
\begin{figure}
\centerline {
\epsfxsize =8.5cm \epsfbox {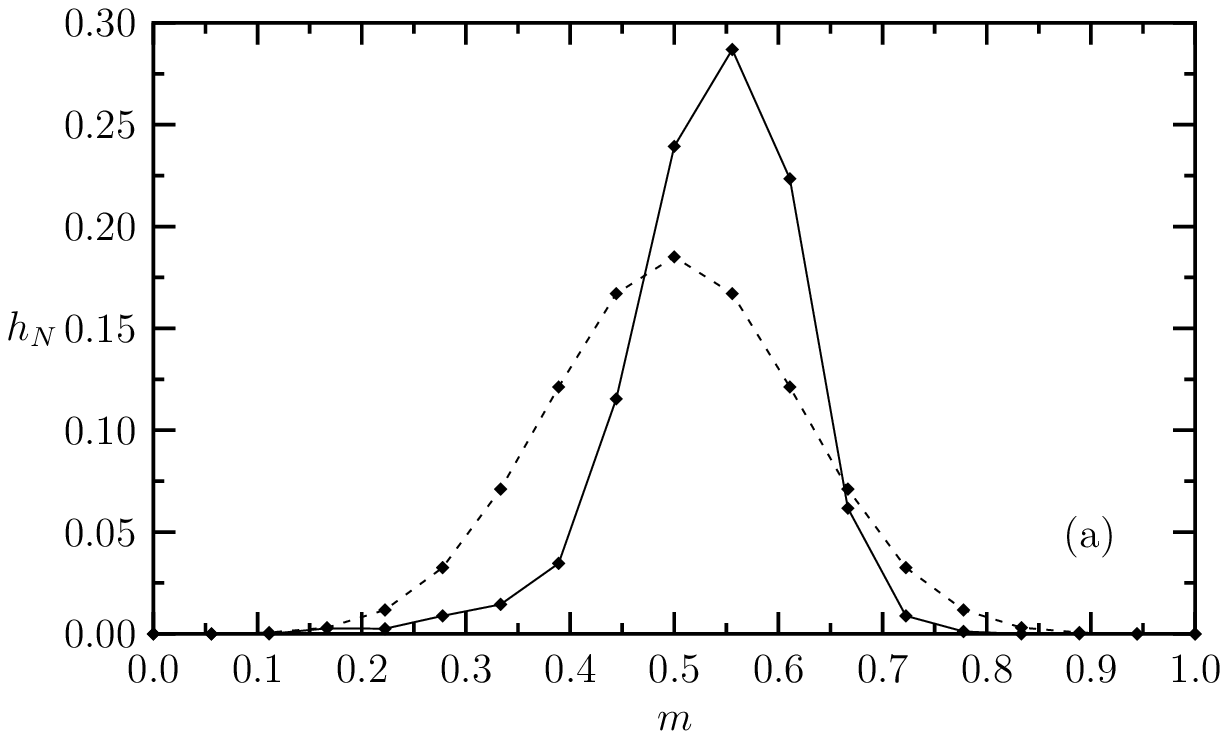}
}

\centerline {
\epsfxsize =8.5cm \epsfbox {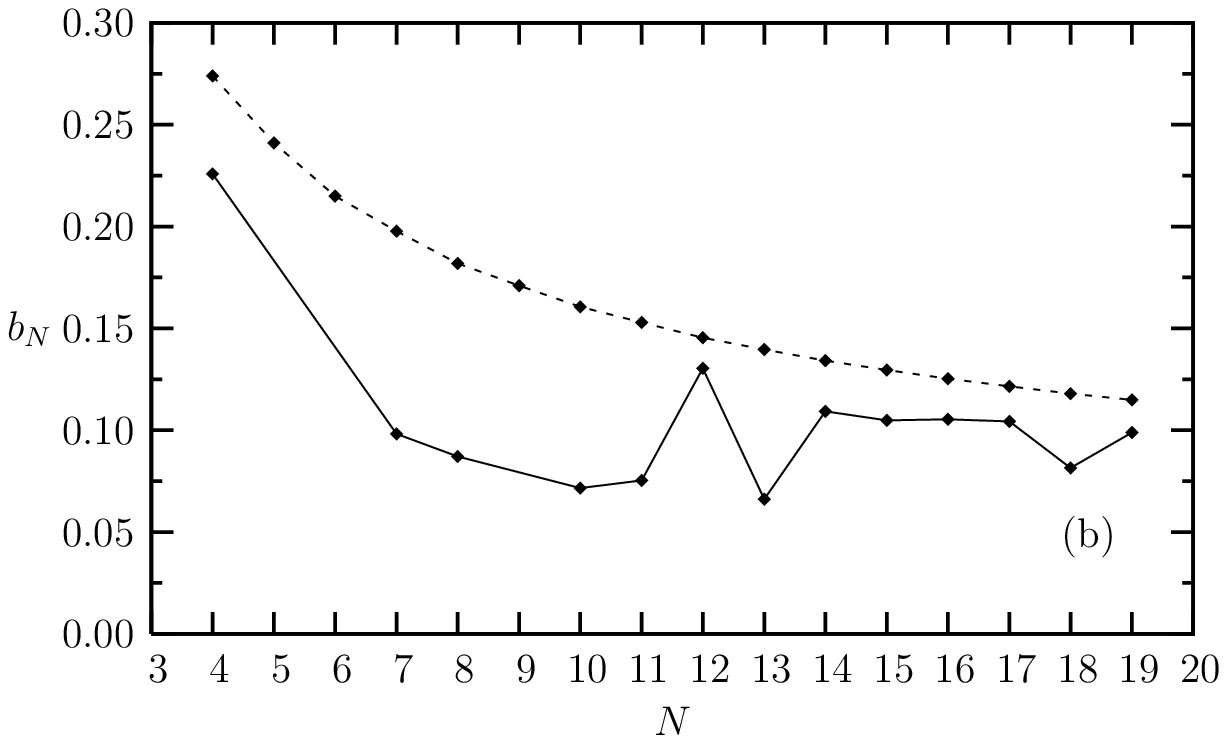}
}
\caption{\label{fighyddist} (a) Distribution of hydrophobicity $h_N$ of all designing sequences with
$N=18$ monomers (solid line) compared with the distribution of hydrophobicity of all sequences
of this length (dashed line) for the MHP model. 
(b) Widths of the hydrophobicity distribution of the designing sequences, $b_N$,
depending on the chain length $N$ (solid line) compared with the widths of the hydrophobicity distribution
of all sequences (dashed line) for the MHP model.}
\end{figure}

In this section, we analyze the complete sets $\mathbf{S}_N$ of designing sequences for HP proteins of given
numbers of residues $N\le 19$. A sequence $\bm{\sigma}$ is 
called {\em designing}, if there is only one conformation
associated with the native ground state, 
not counting rotation, translation, and reflection symmetries
that altogether contribute on a simple cubic lattice a symmetry factor $6$ for linear, $24$ for planar, and $48$ for 
conformations spreading into all three spatial directions. In Table~\ref{tabdesseq} we have listed
the numbers of designing sequences $S_N$ we found for the two models. In contrast to previous
investigations of HP proteins on the square lattice~\cite{irb2}, the number of designing sequences
obtained with the pure HP model is extremely small on the simple cubic lattice. This does not allow 
for a reasonable statistical study of general properties of designing sequences, at least for very short
chains. The situation is much better using the more adequate MHP model.  
The first quantity under consideration is the hydrophobicity of a sequence 
$\bm{\sigma}$, i.e., the number of hydrophobic
monomers $N_H$, normalized with respect to the total number of residues:
\begin{equation} 
\label{hydrop}
m(\bm{\sigma})=\frac{N_H}{N}=\frac{1}{N}\sum\limits_{i=1}^N \sigma_i.
\end{equation}
The average hydrophobicity over a set of designing sequences of given length $N$ is then defined by
\begin{table}
\caption{\label{tabdesseq} Number of designing sequences $S_N$ (only relevant sequences)
in the HP and MHP model}
\centerline{
\begin{tabular}{c|*{16}{c}} \hline \hline
$N$ & 4 & 5 & 6 & 7 & 8 & 9 & 10 & 11 & 12 & 13 & 14 & 15 & 16 & 17 & 18 & 19 \\ \hline
$S^{\rm HP}_N$ & 3 & 0 & 0 & 0 & 2 & 0 & 0 & 0 & 2 & 0 & 1 & 1 & 1 & 8 & 29 & 47 \\
$S^{\rm MHP}_N$ & 7 & 0 & 0 & 6 & 13 & 0 & 11 & 8 & 124 & 14 & 66 & 97 & 486 & 2196 & 9491 & 4885 \\ \hline \hline
\end{tabular}
}
\end{table}
\begin{figure}[t]
\centerline {
\epsfxsize =8.5cm \epsfbox {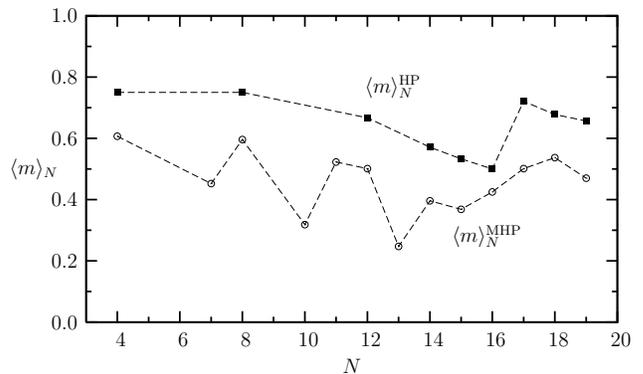}
}
\caption{\label{figavhyd} Dependence of average hydrophobicities $\langle m\rangle_N$ for designing
sequences on the sequence length 
$N$ in both models. For several chain lengths, where no designing sequences exist 
(see Table~\ref{tabdesseq}), the calculation of the average hydrophobicity~(\ref{avhyd}) 
does not make sense. The dashed lines are therefore only guides to the eye.}
\end{figure}
\begin{equation}
\label{avhyd}
\langle m\rangle_N=\frac{1}{S_N}\sum\limits_{\bm{\sigma}\in\mathbf{S}_N} m(\bm{\sigma}).
\end{equation}
In Fig.~\ref{figavhyd} we have plotted $\langle m\rangle_N$ as function of the sequence length $N$. The
plots do not show up a clear tendency to what average hydrophobicity they converge for long chains.
This to know would be, however, of some interest for the design of a biased algorithm of Monte Carlo type
that searches the combined sequence and conformational space for candidates of designing sequences 
with lengths, where enumeration is no longer applicable. A distinct indication that designing sequences 
have in most cases hydrophobicities different from $0.5$
could be used as a bias in order to reduce the section of the sequence space to be 
scanned, since the number of all possible sequences with given
hydrophobicity has a peak at $m=0.5$ (see Fig.~\ref{fighyddist}(a)) which
becomes the more pronounced the higher the number of residues is.
 
It should be noted that the hydrophobicity 
distribution for all these sequences is not binomial since in our analysis we have 
distinguished only sequences that we call {\em relevant}, i.e., two sequences that are
symmetric under reversal of their residues are identified and enter only once into the statistics. 
Therefore we consider, for example, only 10 relevant sequences with
length $N=4$ instead of $2^4=16$. Taking into account {\em all} $2^N$ sequences would obviously
lead to a binomial distribution for $N_H$, since 
there are then exactly 
\begin{equation}
\binom{N}{N_H}
\end{equation}
sequences with $N_H$ hydrophobic monomers.

\begin{figure}
\centerline {
\epsfxsize =8.5cm \epsfbox {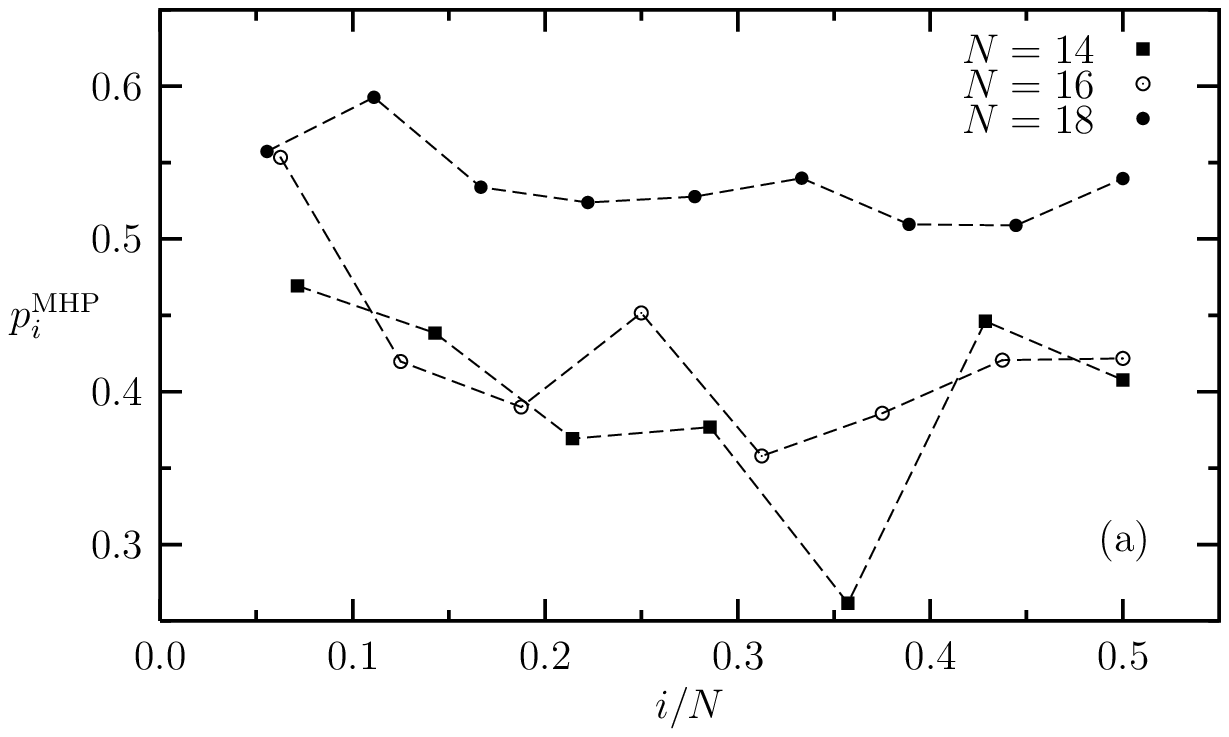}
}

\centerline {
\epsfxsize =8.5cm \epsfbox {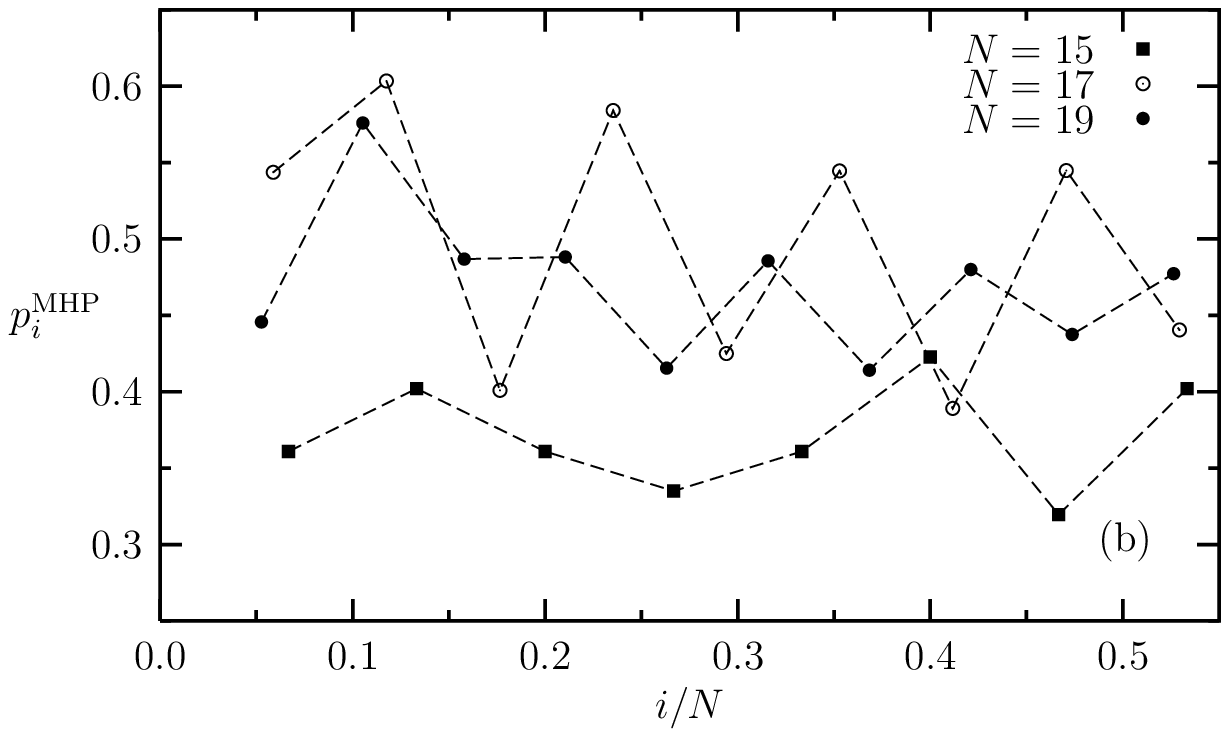}
}
\caption{\label{fighydprof} Hydrophobicity profiles $p_i$ for designing sequences with (a) even and 
(b) odd numbers of monomers in the MHP model. Since the profile (\ref{hydprof}) is symmetric under
$i \leftrightarrow N-i+1 $ we have only plotted it for half the chain.}
\end{figure}

In Fig.~\ref{fighyddist}(a) we have plotted both, the distribution of hydrophobicity of the
designing sequences with $N=18$ monomers in the MHP model and, for comparison, the 
distribution of all sequences with $N=18$. For this example, we see that the width of the 
hydrophobicity distribution for the designing sequences, 
which has its peak at $\langle m\rangle_{18}^{\rm MHP}\approx 0.537 > 0.5$,
is smaller than that of the distribution over all sequences. 
In order to gain more insight how the hydrophobicity distributions differ, 
we have compared the widths of both distributions in their dependence on the chain
length $N\le 19$. This is shown in Fig.~\ref{fighyddist}(b). It seems that for $N\to\infty$ the
widths of the hydrophobicity distributions for the designing sequences asymptotically approach the 
curve of the widths of the hydrophobicity distributions of all sequences.      

\begin{figure}
\centerline {
\epsfxsize =8.5cm \epsfbox {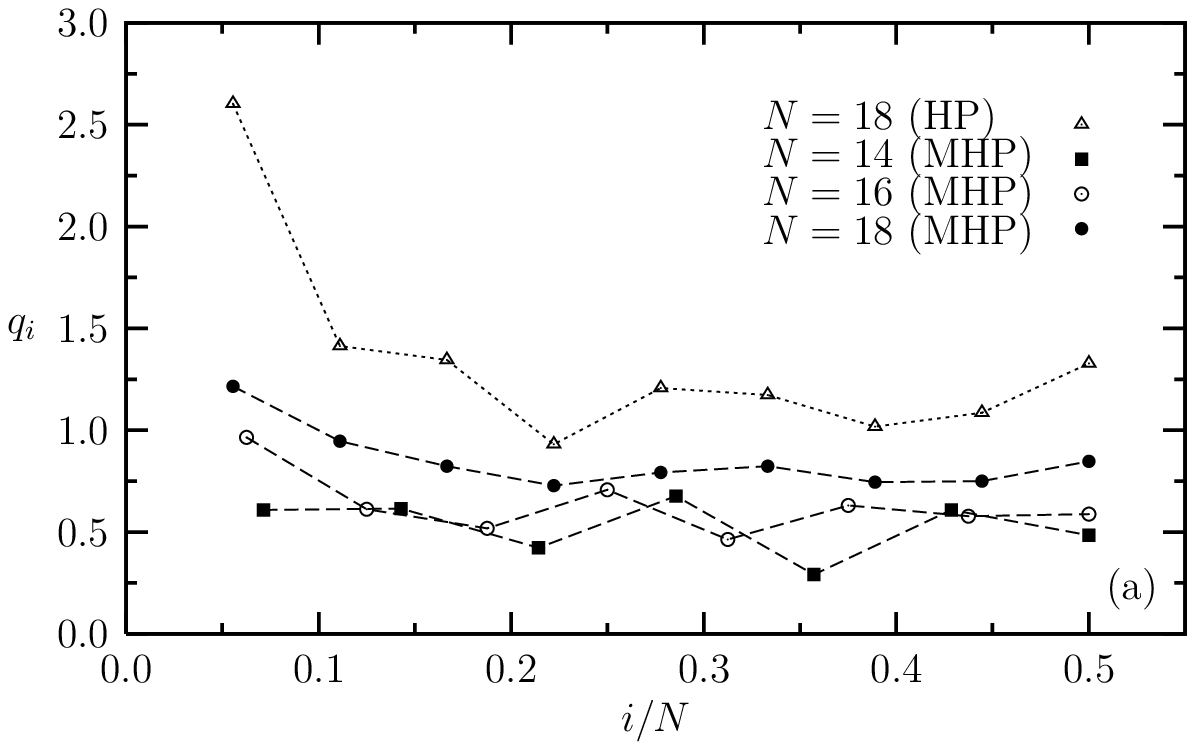}
}

\centerline {
\epsfxsize =8.5cm \epsfbox {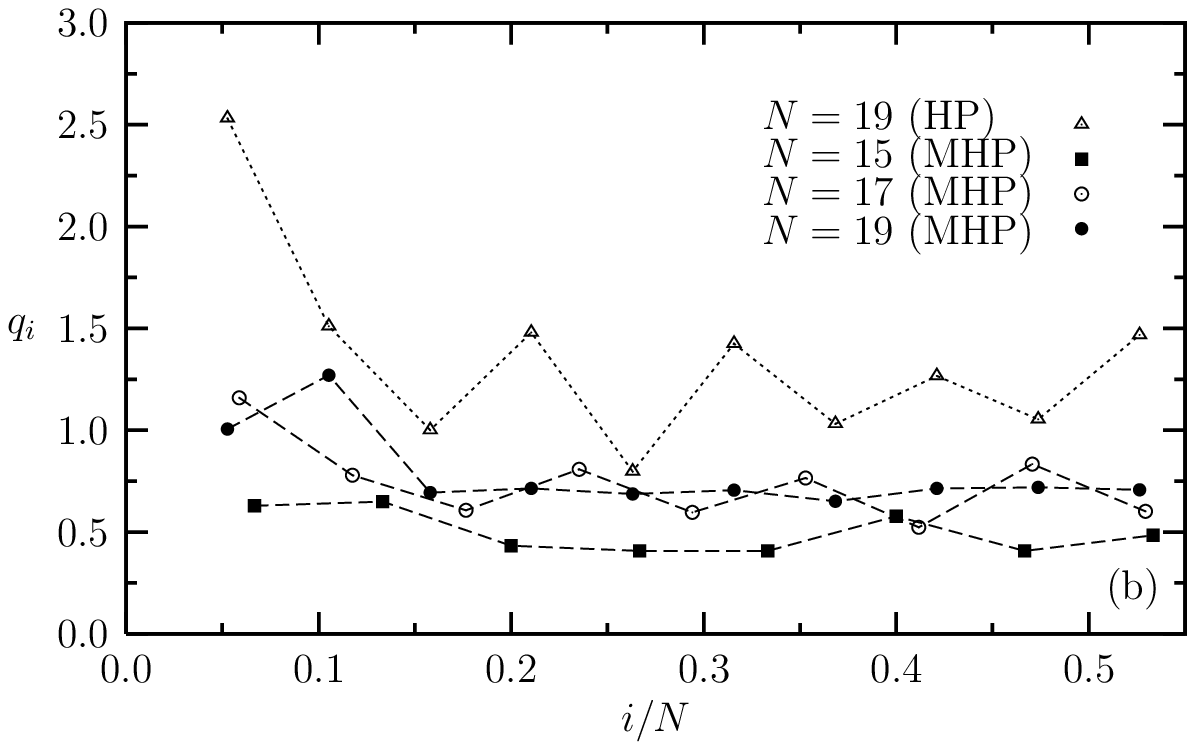}
}
\caption{\label{fighhprof} Profiles of the contact density $q_i$ for designing sequences 
with (a) even and (b) odd numbers of monomers in the MHP model. For comparison, we have 
also inserted the profiles obtained in the HP model for $N=18$ and $N=19$, respectively.
By definition~(\ref{hhprof}), this profile is also symmetric under $i \leftrightarrow N-i+1$.}
\end{figure}

The hydrophobicity profile
\begin{equation}
\label{hydprof}
p_i=\frac{1}{2S_N}\sum\limits_{\bm{\sigma}\in\mathbf{S}_N}(\sigma_i+\sigma_{N-i+1}), \quad i=1,2,\ldots,N,
\end{equation}
is a measure for the probability to find a hydrophobic monomer in a distance $i$ from the nearest end
of a designing sequence. Thus, this quantity gives an impression of how the $H$ monomers are 
on average distributed along the
chain. In Figs.~\ref{fighydprof}(a) and (b) the profiles for designing sequences in the MHP model 
are plotted for respective chains with even ($N=14,16,18$) and odd numbers
($N=15,17,19$) of residues. As a first remarkable result, we see that for odd numbers of monomers 
the profile shows up periodic oscillations, i.e., if the $n$th monomer is preferably hydrophobic the
$(n+1)$th residue is with lower probability. As this effect is stronger for $N=17$ than for $N=19$, 
we expect that the amplitude of these oscillations decreases with increasing number of monomers.
The behavior of the chains with even number of monomers (Fig.~\ref{fighydprof}(a)) is less
spectacular. Here, for increasing number of monomers, the probability seems to become
more and more equally distributed. Therefore, it is more interesting to study how each monomer
of the designing sequences 
is involved in the formation of $HH$ contacts (as well as $HP$ contacts in the MHP model) 
being favored in low-energy conformations. We define the hydrophobic contact density profile
by
\begin{equation}
\label{hhprof}
q_i=\frac{1}{2S_N}\sum\limits_{\bm{\sigma}\in\mathbf{S}_N}\sum\limits_{j=1}^N
\left[C_{ij}\sigma_i\sigma_j+C_{N-i+1\,j}\,\sigma_{N-i+1}\sigma_j\right],
\end{equation}
where $C_{ij}$ is the contact map defined after Eq.~(\ref{hpgen}). The higher the affinity
of the $i$th monomer to form contacts (preferably if it is hydrophobic), the bigger the value 
of $q_i$. This profile is shown in Fig.~\ref{fighhprof} for both models, where we have again
separated even and odd numbers of residues. From the two profiles for the HP model 
($N=18$, $19$), we observe that there is a strong tendency of the monomers at the
ends of the chain ($i=1$, $N$) to form hydrophobic ($HH$) contacts. The reason is 
that these two monomers can have 5 nearest neighbors on the s.c.\ lattice, 
i.e., there is one more possibility
for them to form a favorable energetic contact than for monomers residing within the chain.
In the MHP model, the behavior is less unique, since also $HP$ contacts are attractive
and the tendency that the ends are preferably hydrophobic is much weaker. 

\begin{table}[t]
\caption{\label{tabdesign} Number of designable conformations $D_N$ in both models.}
\centerline{
\begin{tabular}{c|*{16}{c}} \hline \hline
$N$ & 4 & 5 & 6 & 7 & 8 & 9 & 10 & 11 & 12 & 13 & 14 & 15 & 16 & 17 & 18 & 19 \\ \hline
$D^{\rm HP}_N$ & 1 & 0 & 0 & 0 & 2 & 0 & 0 & 0 & 2 & 0 & 1 & 1 & 1 & 8 & 28 & 42 \\ 
$D^{\rm MHP}_N$ & 1 & 0 & 0 & 2 & 2 & 0 & 5 & 6 & 30 & 8 & 31 & 58 & 258 & 708 & 1447 & 1623 \\ \hline \hline
\end{tabular}
}
\end{table}

After having discussed sequential properties of designing sequences, we now 
analyze the properties of their unique ground-state structures, the native conformations. 
From Table~\ref{tabdesign} we
read off that the number of {\em different} native conformations $D_N$ is usually much smaller 
than the number of designing sequences, i.e., several designing sequences share the
same ground-state conformation. The number of
designing sequences that fold into a certain given target conformation 
${\bf X}^{(0)}$ (or conformations being trivially symmetric to this by translations, 
rotations, and reflections) is called {\em designability}~\cite{tang4}:
\begin{equation}
\label{design}
F_N({\bf X}^{(0)})=\sum\limits_{\bm{\sigma}\in\mathbf{S}_N}
\Delta\left({\bf X}_{\rm gs}(\bm{\sigma})-{\bf X}^{(0)}\right),
\end{equation}
where ${\bf X}_{\rm gs}(\bm{\sigma})$ is the native (ground-state) 
conformation of the designing sequence
$\bm{\sigma}$. The function $\Delta({\bf Z})$ is the generalization of Eq.~(\ref{defDelta})
to $3N$-dimensional vectors. It is unity for ${\bf Z}={\bf 0}$ and zero otherwise.
\begin{figure}
\centerline {
\epsfxsize =8.5cm \epsfbox {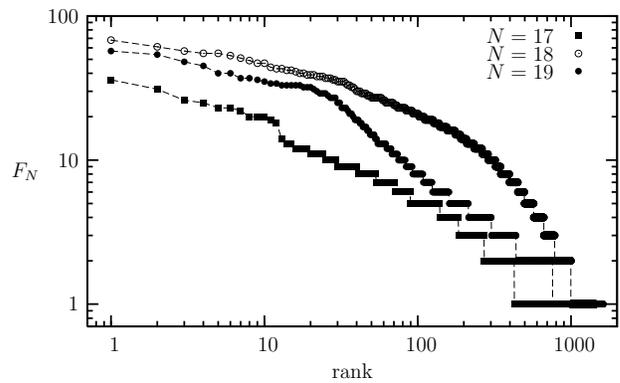}
}
\caption{\label{figdesign} 
Designability $F_N$ of native conformations 
in the MHP model for $N=17$, $18$, and $19$. The abscissa is the rank obtained 
by ordering all designable conformations according to their designability.
}
\end{figure}

\begin{figure}[b]
\centerline{
\parbox{4.4cm}{
\epsfxsize =4cm \epsfbox {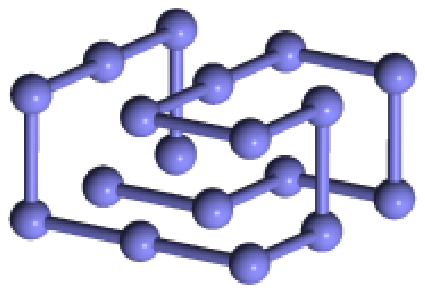}}
\hfill
\parbox{4.4cm}{
\epsfxsize =4cm \epsfbox {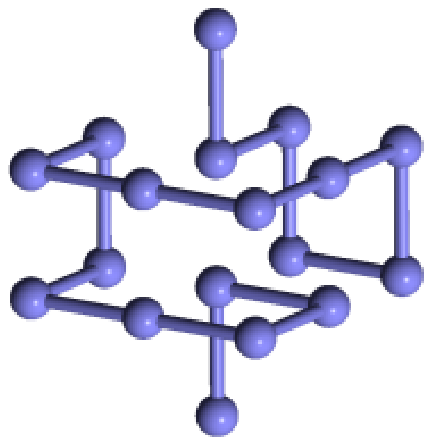}}
}
\caption{\label{fighighmax} 
Structure ($N=18$) with the highest designability of all native 
conformations (left) and with minimal radius of gyration (right).
}
\end{figure}

The designability 
is plotted in Fig.~\ref{figdesign} for all native
conformations that HP proteins with $N=17$, $18$, and $19$ monomers can form in the MHP
model. In this figure, 
the abscissa is the rank of the conformations, ordered according to their 
designability. 
The conformation with the lowest rank is therefore the most designable structure
and we see that a majority of the designing sequences folds into a few number of 
highly designable conformations, while only a small number of designing sequences 
possesses a native conformation with low designability (note that the plot is logarithmic). 
Similar results were found, for example in Ref.~\cite{tang3}, where the designability of
compact conformations on cuboid lattices was investigated in detail. The left picture
in Fig.~\ref{fighighmax} shows the conformation with the lowest rank (or highest designability) 
with $N=18$ monomers.

From our analysis we see that this characteristic
distribution of the designing sequences is not restricted to cuboid lattices only.
This result is less trivial than one may think at first sight. As we will show later on in this paper
in the discussion of the radius of gyration, native conformations are very compact, but
only very few conformations are maximally compact (at least for $N\le 19$). For longer sequences 
similar results were found in Ref.~\cite{bj2}.  
Highly designable conformations are of great interest, since it is expected that they form 
a frame making them stable against mutations and thermodynamic fluctuations. Such
fundamental structures are also relevant in nature, where in particular secondary 
structures (helices, sheets, hairpins) supply proteins with a stable backbone~\cite{tang3}. 

Conformational properties of polymers are usually studied in terms of
the squared end-to-end distance
\begin{equation}
\label{sqeed}
R_e^2=({\bf x}_N-{\bf x}_1)^2
\end{equation}
and the squared radius of gyration
\begin{equation}
\label{sqrog}
R_g^2=\frac{1}{N}\sum\limits_{i=1}^N({\bf x}_i-\bar{\bf x})^2,
\end{equation}
where $\bar{\bf x}=\sum_i {\bf x}_i/N$ is the center of mass of the polymer.
\begin{figure}
\centerline {
\epsfxsize =8.5cm \epsfbox {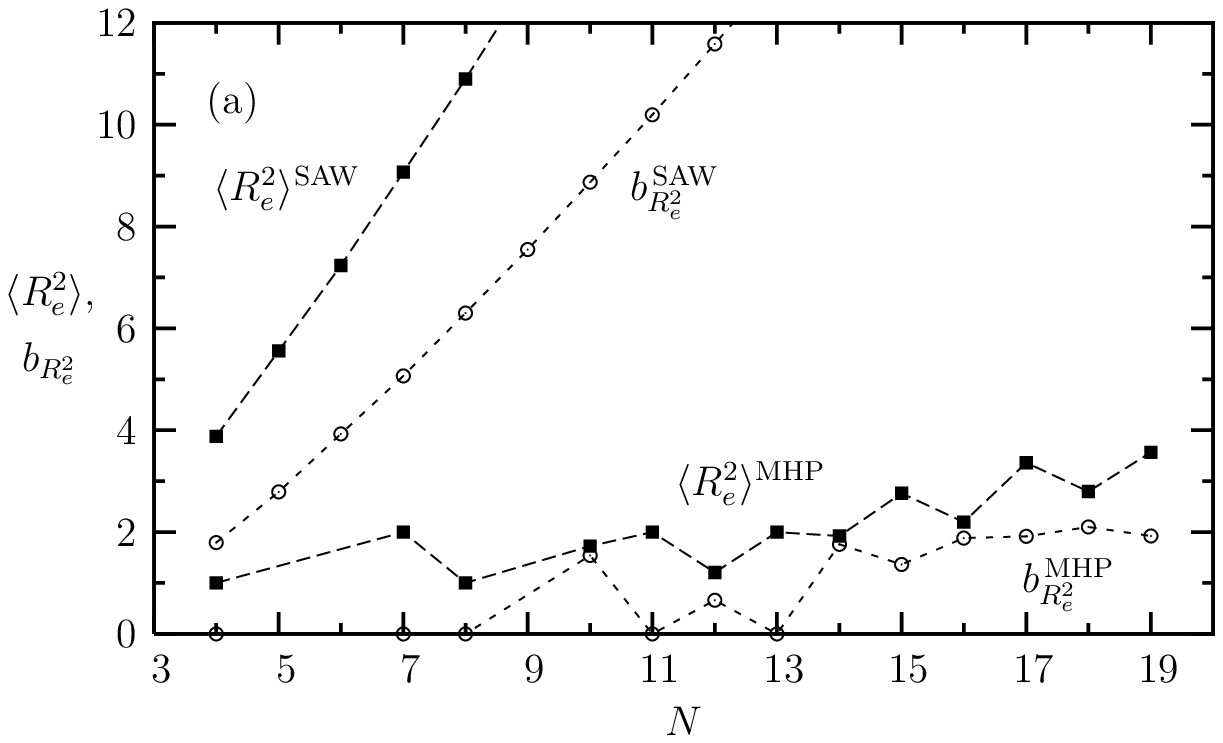}
}

\centerline {
\epsfxsize =8.5cm \epsfbox {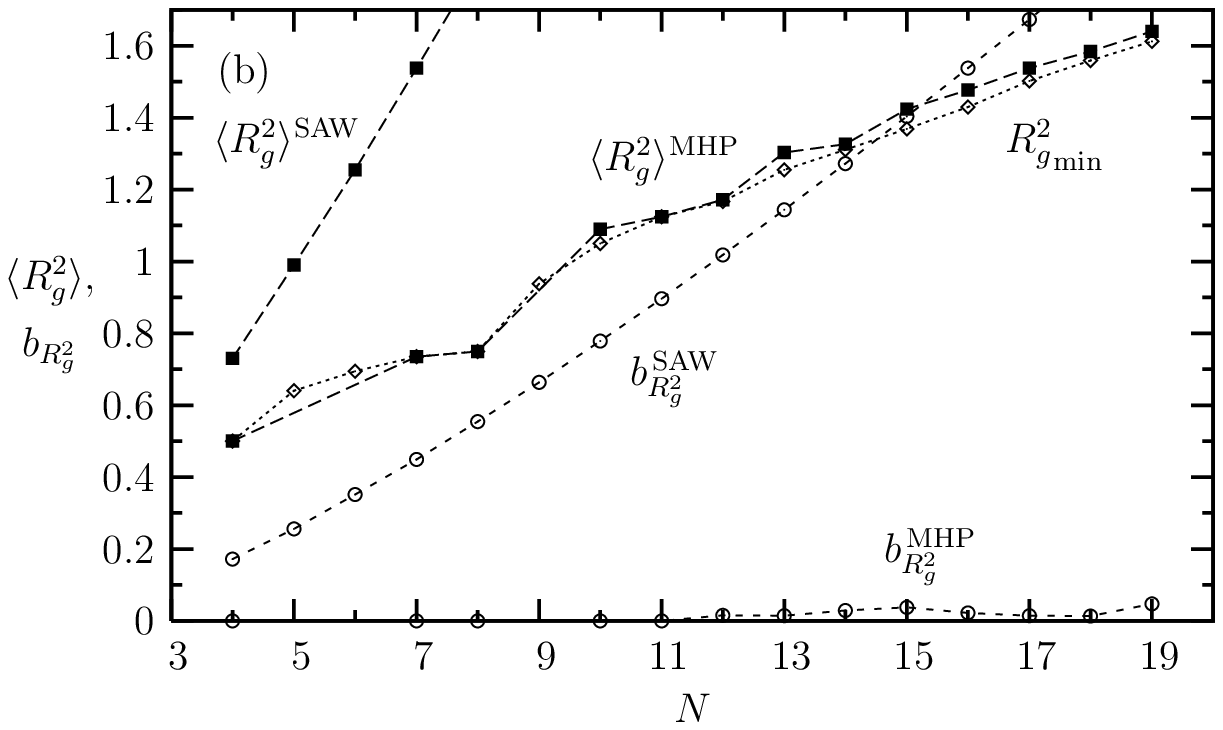}
}
\caption{\label{figavgeo} (a) Average squared end-to-end distances $\langle R_e^2\rangle$ 
of native conformations in the MHP model compared with those of all self-avoiding 
walks (SAW). We have also inserted the widths $b_{R_e^2}$ of the corresponding 
distributions of end-to-end distances.
(b) The same for the average squared radius of gyration $\langle R_g^2\rangle$. 
Since the radius of gyration is an appropriate measure for
the compactness of a conformation, we have also plotted ${R^2_g}_{\rm min}$ 
for the conformations with the minimal radius of gyration (or, equivalently, 
maximal compactness). 
}
\end{figure}

In polymer physics both quantities are usually referred to as measures for
the compactness of a conformation. In Fig.~\ref{figavgeo}(a) we compare the $N$-dependence
of the averages of the native conformations found in the MHP model and all
possible self-avoiding walks. The same quantities for the squared radius of gyration 
are shown in Fig.~\ref{figavgeo}(b). The averages were obtained by calculating
\begin{eqnarray}
\label{aveedSAW}
\langle R_{e,g}^2\rangle^{\rm SAW} &=& \frac{1}{C_N}\sum\limits_{{\bf X}\in \mathbf{C}_N} 
R_{e,g}^2({\bf X}),\\
 \label{aveedMHP}
\langle R_{e,g}^2\rangle^{\rm MHP} &=& 
\frac{1}{S_N}\sum\limits_{\bm{\sigma}\in\mathbf{S}_N} R_{e,g}^2({\bf X}_{\rm gs}(\bm{\sigma})),
\end{eqnarray}
where $\mathbf{C}_N$ is the set of all self-avoiding conformations on a s.c.\ lattice.
Figure~\ref{figavgeo}(a) 
shows that, compared to $\langle R_{e,g}^2\rangle^{\rm SAW}\sim n^{2\nu}$ with 
$\nu\approx 0.59$ (see Ref.~\cite{wong1} for a recent summary of estimates for $\nu$), 
the average 
end-to-end distance $\langle R_{e,g}^2\rangle^{\rm MHP}$ for the native conformations 
only is much smaller. For even number of monomers, the ends of a HP protein can form
contacts with each other on the s.c.\ lattice. Accordingly, the values of 
$\langle R_{e,g}^2\rangle^{\rm MHP}$ are smaller for $N$ being even and
the even-odd oscillations are very pronounced. The widths (or standard deviations) 
$b_{R_e^2}$ of the distributions of the squared end-to-end distances are also 
very small. Even for heteropolymers with $N=19$ monomers in total, there are virtually no 
native conformations, where the distance between the ends is larger than 3 lattice sites.
We have checked this for the HP model, too, and found the same effect. Since the number 
of native conformations is very small in this model, we have not included these results 
in the figure. Depicting the average squared radius of gyration $\langle R^2_g\rangle$
and the widths of the corresponding distribution of the radius of gyration in 
Fig.~\ref{figavgeo}(b) for all self-avoiding conformations as well as for the native ones, 
we see that these results confirm the above remarks. As the average 
end-to-end distances of native conformations are much smaller than those for the
bulk of all conformations, we observe the same trend for the mean squared 
radii of gyration $\langle R_g^2\rangle^{\rm MHP}$ and 
$\langle R_g^2\rangle^{\rm SAW}$ and the widths $b_{R_g^2}^{\rm \,MHP}$ and
$b_{R_g^2}^{\rm \,SAW}$ as well. In particular, the width $b_{R_g^2}^{\rm \,MHP}$
is so small, that virtually all native conformations possess the same 
radius of gyration. For this reason, we have also searched for the conformations
having the smallest radius of gyration ${R_g^2}_{\rm min}$ (these conformations 
are not necessarily native as we will see!) and inserted these values into this 
figure, too. We observe that these values differ only slightly from 
$\langle R_g^2\rangle^{\rm MHP}$. Thus we conclude that native conformations
are very compact, but not necessarily maximally compact. This property has already been 
utilized in enumerations being performed {\em a priori} on compact
lattices~\cite{tang1,tang2,tang3}, where, however, the proteins are confined by hand 
to live in small cuboids (e.g.\ of size $3\times3\times3$ or $4\times3\times3$).
Our results on the general s.c.\ lattice confirm that this assumption is
justified to a great extent. Nevertheless, the slight deviation
from the minimal radius of gyration native conformations exhibit is a remarkable
result as it concerns about $90\%$ of the whole set of native conformations!
\begin{figure}
\centerline {
\epsfxsize =8.5cm \epsfbox {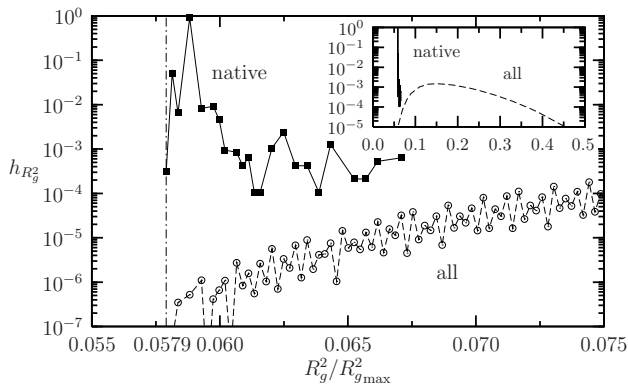}
}
\caption{\label{fighistgyr} Distribution $h_{R_g^2}$ (normalized 
to $\sum h_{R_g^2} = 1$) of squared radii of gyration  
(normalized with respect to the maximal radius of gyration ${R_g^2}_{\rm max} = (N^2-1)/12$ 
of a completely stretched conformation) of native conformations with $N=18$ in the MHP model, 
compared with the histogram for all self-avoiding conformations. The vertical line
refers to the minimal radius of gyration (${R_g^2}_{\rm min}/{R_g^2}_{\rm max}=0.0579$
for $N=18$) and an associated structure is shown on the 
right-hand side of Fig.~\ref{fighighmax}. The inset shows the distribution
up to $R_g^2/{R_g^2}_{\rm max}=0.5$.  
}
\end{figure}
This can be seen in Fig.~\ref{fighistgyr}, where we have plotted the distribution of 
the squared radii of gyration for all self-avoiding conformations with $N=18$ and the
native states in the MHP model. All native conformations have a very small radius of gyration
but only a few of them share the smallest possible value. A structure with the smallest
radius of gyration is shown on the right-hand side of Fig.~\ref{fighighmax}. It obviously
differs from the most-designable conformation drawn on the left of the same figure. 
\section{Density of States and Thermodynamics} 
\label{sectdens}
In this section we systematically compare energetic thermodynamic quantities
of designing and non-designing sequences. In Ref.~\cite{bj3} it was conjectured 
for exemplified sequences of comparable 14mers, one of them being designing, that
designing sequences in the HP model seem to show up a much more pronounced low-temperature
peak in the specific heat than the non-designing examples. This peak may be
interpreted as kind of a conformational transition between structures with
compact hydrophobic cores (ground states) and states where the whole conformation 
is highly compact (globules)~\cite{bj1,bj2}. Another peak in the specific heat
at higher temperatures, which is exhibited by all lattice proteins, is an indication
for the usual globule -- coil transition between compact and untangled conformations.

In order to study energetic thermodynamic quantities such as mean energy and
specific heat we determined from our enumerated conformations for a given sequence 
the density of states $g(E)$ that conveniently allows the calculation of the
partition sum $Z(T)=\sum_E g(E)\exp(-E/k_BT)$ and the moments 
$\langle E^k\rangle_T=\sum_E E^k g(E)\exp(-E/k_BT)/Z$, where the subscript
$T$ indicates the difference of calculating thermal mean values based
on the Boltzmann probability from averages previously introduced
in this paper. Then the specific heat is given
by the fluctuation formula $C_V=(\langle E^2\rangle_T-\langle E\rangle_T^2)/k_BT^2$.
\subsection{Sequences in the HP Model}
\label{subsectHP}
\begin{figure}[t]
\centerline {
\epsfxsize =8.5cm \epsfbox {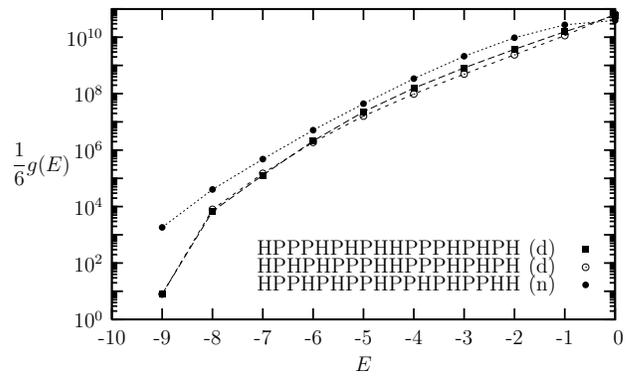}
}
\caption{\label{figdos_HP} Density of states $g(E)$ for two designing sequences (d) with 
$N=18$, $m_H=8$, and $E_{\rm min}=-9$ in the HP model. We have divided out the symmetry 
factor $6$ that is common to all conformations. Three-dimensional conformations
have an additional symmetry factor $8$, such that the states with minimal energy
for these two curves are indeed unique and the sequences are designing. For comparison we 
have also plotted $g(E)$ for one exemplified non-designing sequence (n) out of $525$ 
having the same properties as quoted above, but different sequences. The ground-state
degeneracy for this example is $g_0=g(E_{\rm min})=6\times 1840$ 
(including all symmetries). }
\end{figure}
\begin{figure}
\centerline {
\epsfxsize =8.5cm \epsfbox {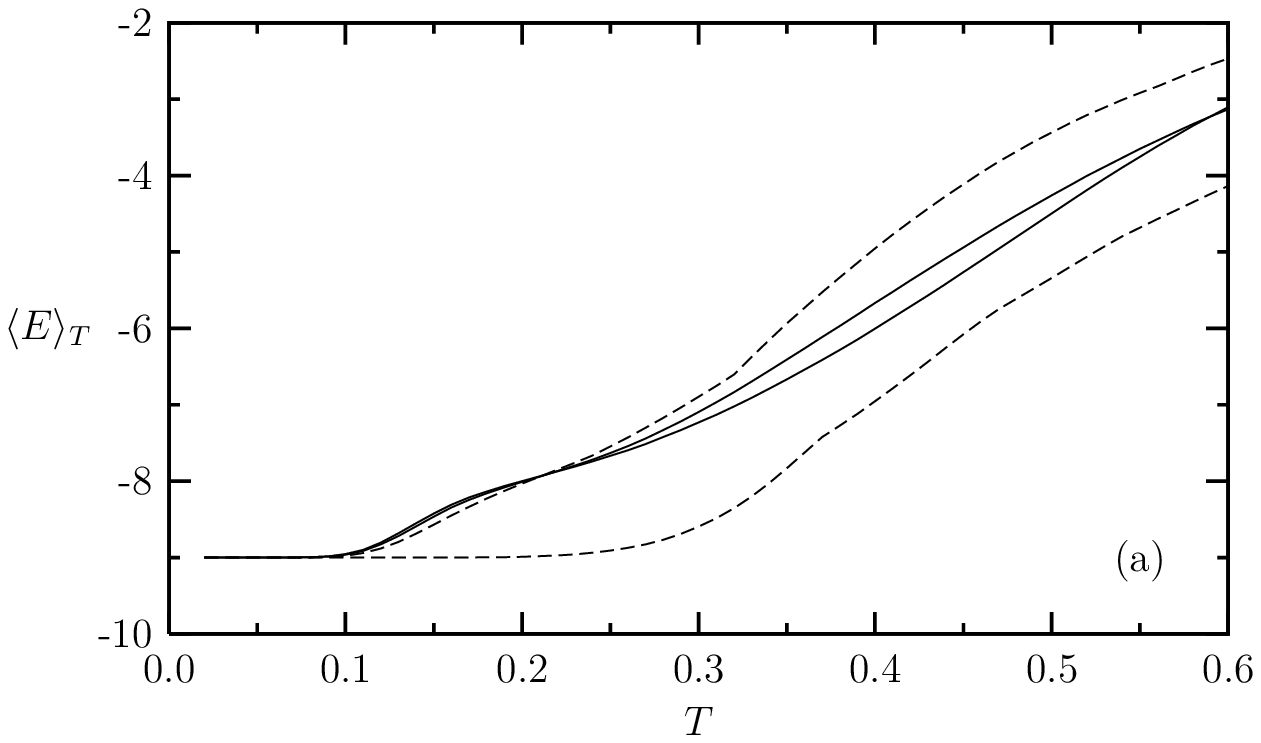}
}

\centerline {
\epsfxsize =8.5cm \epsfbox {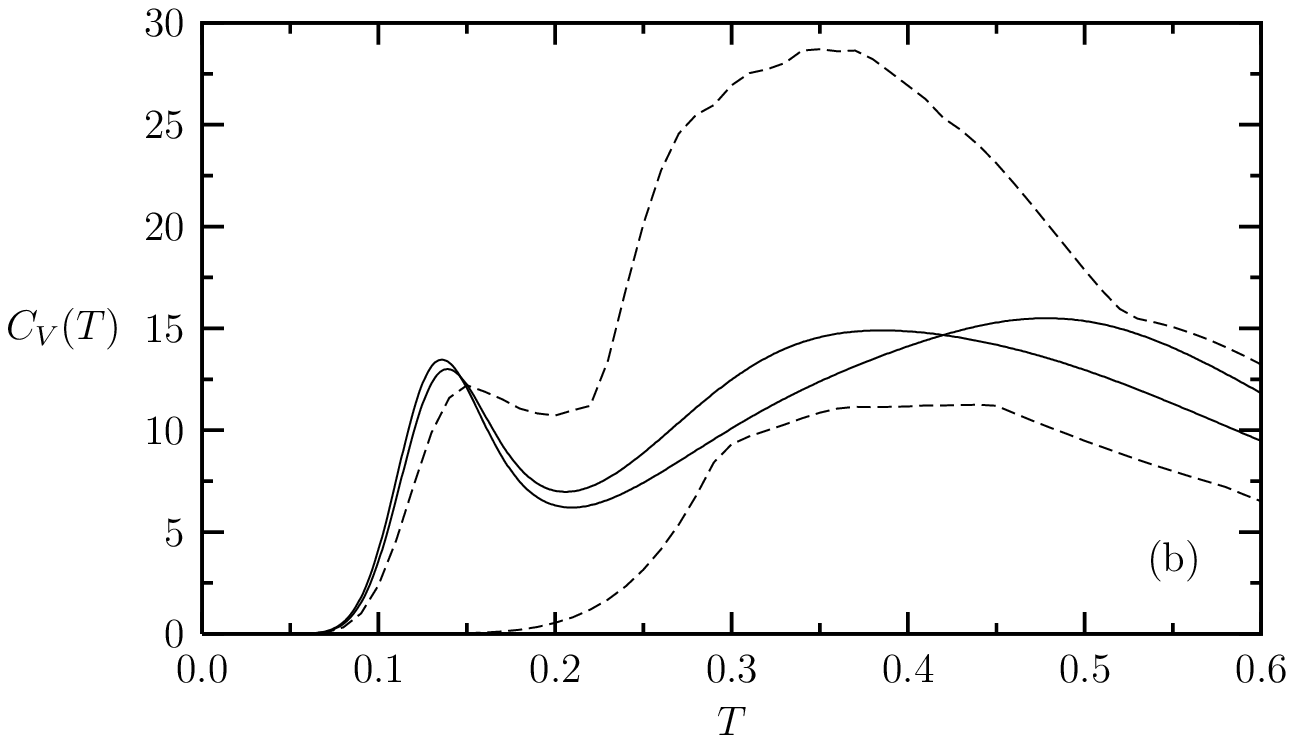}
}
\caption{\label{figEC_HP} (a) Mean energy $\langle E\rangle_T$ and (b) specific heat $C_V$
for the two designing sequences with $N=18$, $m_H=8$, and $E_{\rm min}=-9$ (solid lines) in the HP model, 
whose densities of states were 
plotted in Fig.~\ref{figdos_HP}. The curves of the same quantities for the 
$525$ non-designing sequences 
are completely included within the respective areas between the dashed lines. 
The low-temperature peak
of the specific heat (near $T=0.14$) is most pronounced for the two designing sequences which
behave similarly for low temperatures. 
}
\end{figure}
In the HP model with pure hydrophobic interaction the density of states shows up
a monotonic growth with increasing energy, at least for the short chains 
in our study (for longer chains, e.g.\ the 42mer investigated in Ref.~\cite{bj1,bj2},
the number of states in the high-energy region decreases, i.e., the density of states 
possesses a global maximum at an energy $E$ residing within the interval 
$E_{\rm min} < E < E_{\rm max}=0$). For a reasonable comparison of the behavior of
designing and non-designing sequences, we have focused on 18mers having the same
hydrophobicity ($m_H=8)$ and ground-state energy $E_{\rm min}=-9$. There are in total  
$527$ sequences with these properties, two of which are designing. 
The densities
of states for the two designing sequences and an example of a non-designing sequence 
are plotted in Fig.~\ref{figdos_HP}. We have already divided out a global symmetry
factor $6$ (number of possible directions for the link connecting the first two monomers)
that all conformations on a s.c.\ lattice have in common. Since the
ground-state conformations of the designing sequences spread into all three
dimensions, an additional symmetry factor $4\times 2=8$ ($4$ for rotations around the first bond,
$2$ for a remaining independent reflection) makes a number of conformations
obsolete and the ground-state degeneracy of the designing sequences is indeed unity.
Obviously this is not the case for the sequences we identified as non-designing.
In fact, the uniqueness of the ground states of designing sequences is a remarkable property 
as there are not less than $\sim 10^{10}$ possible conformations of HP lattice
proteins with 18 monomers.
As we also see in Fig.~\ref{figdos_HP}, the ratio of the density of the first excited 
state ($E=-8$) for the designing and the non-designing sequences 
is smaller than for the ground state. This means that, at least for these short
chains, the low-temperature behavior of the HP proteins in this model strongly
depends on the degeneracy of the ground state. Furthermore, we expect that the 
low-temperature behavior of both designing sequences is very similar as their
low-energy densities hardly differ. We have investigated this, once more
for the 18mers with the properties described above, by considering 
the mean energy $\langle E\rangle_T$ and the specific heat $C_V(T)$. The results
are shown in Figs.~\ref{figEC_HP}(a) and (b), respectively. 
The two solid curves belong to the two designing sequences and the dashed lines are the
minimum/maximum bounds of the respective quantities for the non-designing sequences. 
As a main result we find that designing and non-designing sequences behave indeed 
differently for very low temperatures. There is a characteristic, pronounced 
low-temperature peak in the specific heat that can be interpreted as kind
of transition between low-energy states with hydrophobic core and 
very compact globules. This confirms a similar observation for the 14mers
studied in Ref.~\cite{bj3}. 

The upper bound of the specific heats for non-designing 
sequences in Fig.~\ref{figEC_HP}(b) exposes two peaks. 
By analyzing our data for all $525$ non-designing sequences
we found that there are two groups: some of them
experience two conformational transitions, while others do not show 
a characteristic low-temperature behavior. Thus, the only appearance of 
these two peaks is not a unique, characteristic
property of designing sequences. In order to quantify this observation, we have studied all relevant 
$32896$ sequences with 16 monomers. Only one of these sequences is designing ($HP_2HP_2HPHPH_2PHPH$,
with minimum energy $E_{\rm min}=-9$), but in total there are 593 sequences, i.e., $1.8\%$ of
the relevant sequences, corresponding to curves of specific heats with two local maxima.
It should be noted that the degeneracies of the ground states associated with these
sequences are comparably small. 

\begin{figure}[t]
\centerline{
\epsfxsize = 8.5cm \epsfbox{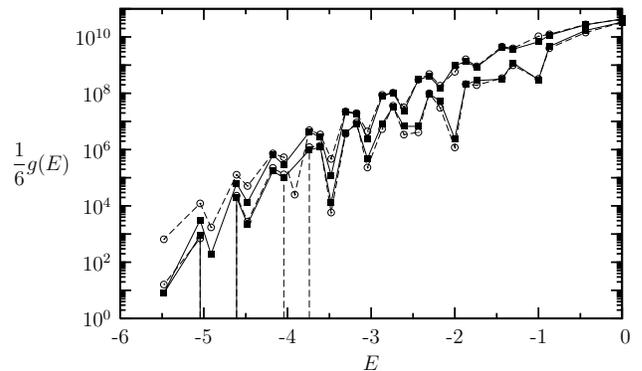}
}
\caption{\label{figdos_MHP}
Minimum and maximum boundaries for the densities of states of the 13 designing (filled boxes, connected
by solid lines)
and the 40 non-designing sequences (open circles, connected by dashed lines) with 18 monomers, hydrophobicity $m_H=3$, and
ground-state energy $E_{\rm min}\approx -5.478$ in the MHP model. Once more, a global symmetry factor 6 has already been
divided out. 
}
\end{figure}
\subsection{Properties of the MHP Model}
\label{subsectMHP}
\begin{figure}
\centerline{
\epsfxsize = 8.5cm \epsfbox{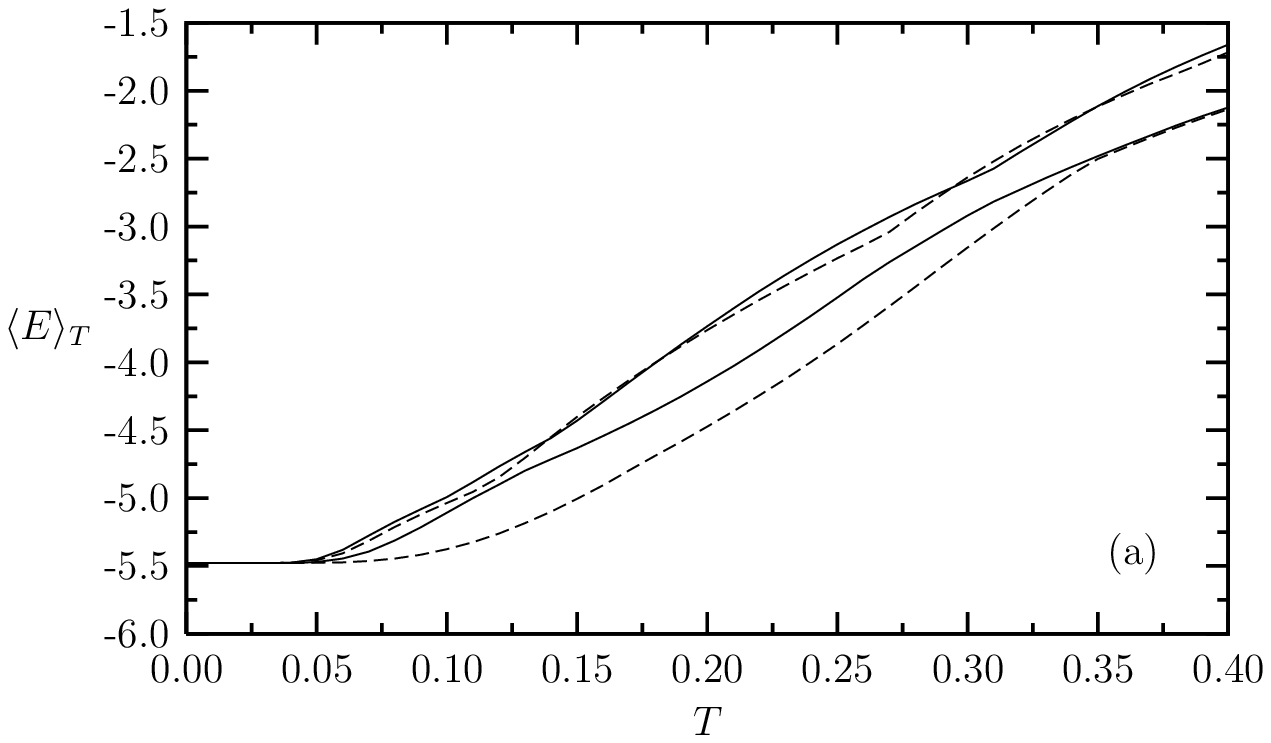}
}

\centerline{
\epsfxsize = 8.5cm \epsfbox{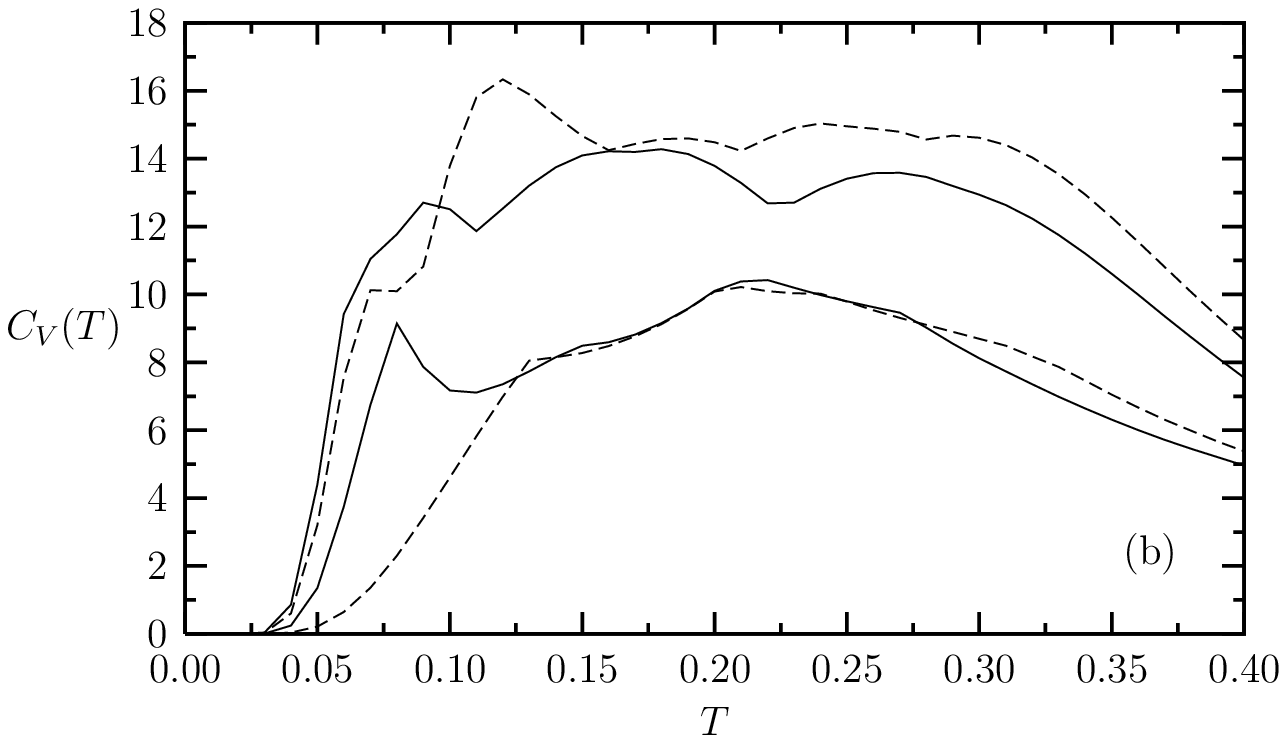}
}
\caption{\label{figECMHP}
Minimum and maximum boundaries of the (a) mean energy $\langle E\rangle_T$ and (b) specific heat $C_V$ for  
the designing (solid lines) and non-designing sequences (dashed lines) in the MHP model, 
with the properties listed in the caption of Fig.~\ref{figdos_MHP}.
}
\end{figure}
In the MHP model, the energy levels are no longer equally spaced due to the additional 
non-integer $HP$ interaction. Moreover, the absolute value of the energy of the lattice 
heteropolymer is not necessarily identical with the number of hydrophobic contacts. The formation 
of a highly compact core is still desirable, but also the attractive contacts between $H$ and $P$
monomers reduce the energy of the heteropolymer. For this reason, the relatively manifest
distinction between ``phases'' with compact $H$-core states and entirely compact conformations
is expected to be much less pronounced, even for the designing sequences.  

Once more, we have first enumerated the densities of states for a set of sequences that
have similar properties but differ only in the ordering of the sequence. For this study
we chose all 18mers sharing the same hydrophobicity $m_H=3$ and identical ground-state energy
$E_{\rm min}=2u_{HH}^{\rm MHP}+8u_{HP}^{\rm MHP}\approx -5.478$, since there are 2 $HH$ 
and 8 $HP$ contacts in each of the ground-state conformations.
We found 13 designing and 40 non-designing sequences that 
satisfy these specifications. In order not to be confused by too many curves, in 
Fig.~\ref{figdos_MHP} again only the minimum and maximum boundaries of the designing 
sequences (solid lines) are shown as well as the corresponding bounds for the
non-designing sequences (dashed lines). We observe that the regions enclosed by the boundaries
do not exhibit significant differences in both cases except for the ground-state
energy $E_{\rm min}$, where the ground-state degeneracy for the designing sequences is
$g_0^{\rm d}={g_0^{\rm d}}_{\rm min}={g_0^{\rm d}}_{\rm max}=48$ (i.e., identical with the global 
symmetry factor for 3D conformations), while the degeneracies for the non-designing sequences
lie within the interval ${g_0^{\rm n}}_{\rm min}=96 \le g_0^{\rm n} \le {g_0^{\rm n}}_{\rm max}=3888$. 
Note that $g^{\rm n}_{\rm min}(E\approx -3.913)=0$ and 
$g^{\rm d}_{\rm min}(E\approx -4.913)=g^{\rm n}_{\rm min}(E\approx -4.913)=0$. Interestingly, the state
with energy $E\approx -3.913$ is never occupied by the designing sequences.

Since the densities of states for designing and non-designing sequences hardly differ,
it is difficult to identify a particular thermodynamic behavior being characteristic for designing
sequences only. This is indeed true as can be seen from Figs.~\ref{figECMHP}(a) and (b), where we
have plotted the lower and upper boundaries for the respective mean energies and specific heats of these 
18mers. In contrast to the results for the HP model (cf.\ Figs.~\ref{figEC_HP}(a) and (b)), where,
within a certain low-temperature interval, the regions enclosing the curves for the designing and
non-designing sequences do not overlap, a separation of this kind is not apparent in the MHP model.
Nevertheless, these figures also show that for very low temperatures ($0 < T < 0.1$), the general
behavior is very similar for all designing sequences, but it is not for the non-designing sequences,
where the temperature dependence of energy and thus specific heat can be significantly different. 
\section{Summary}
\label{sum}
We have exactly analyzed the combined space of sequences and conformations 
for proteins on the simple cubic lattice for two HP-type models that 
differ in the contact energy
between hydrophobic and polar monomers. In the original HP model~\cite{dill2}
this interaction is zero, while in the more realistic MHP model~\cite{tang1} 
there is a nonzero contribution as suggested by the Miyazawa-Jernigan matrix
of contact energies between amino acids in proteins~\cite{mj1}. Since there
were only a few known exact results for heteropolymers in 3D, in particular
on compact cuboid lattices, we generated by exact enumeration the sets of 
designing sequences and native conformations on non-compact simple cubic
lattices. We studied, how their properties,
measured, e.g., in terms of quantities like end-to-end distance, radius of gyration,
designability, etc., differ from the bulk of all possible sequences and 
all self-avoiding conformations, respectively. We found that ground states of 
designing sequences, i.e.\ native conformations, have a much smaller 
mean end-to-end distance than the set of all conformations with the same length. Moreover,
we confirmed that these conformations are very compact, i.e., they have a smaller
mean radius of gyration than the whole set. This is valid for both models under
consideration. 

We have also studied energetic thermodynamic properties, in order to investigate
how characteristic the low-temperature behavior of designing compared to non-designing 
sequences is. We determined the densities of states for respective sets of selected 18mers with
comparable properties. In the HP model, where the number of designing sequences
is rather small compared with the MHP model, we could observe that energetic
fluctuations are different for designing and non-designing sequences within a certain
low-temperature region. Designing sequences show up a more pronounced low-temperature
peak in the specific heat being related to a conformational transition between 
low-energy states with hydrophobic core and highly compact globules. For the MHP model
the situation is more diffuse, and a clear distinction between designing and non-designing sequences
based on characteristic thermodynamic properties is not uniquely possible. Nevertheless,
we have also seen in this model that designing sequences behave similarly for very low
temperature while non-designing sequences react quite differently on changes
of the temperature, over the entire range of temperatures.

\section{Acknowledgements}
This work is partially supported by the German-Israel-Foundation (GIF) under
contract No.\ I-653-181.14/1999. One of us (R.S.) acknowledges support by the
\mbox{Studienstiftung des deutschen Volkes}.

\begin{thebibliography}{199}
\bibitem{fnscale}
In contrast to Ref.~\cite{tang1}, we have kept $u_{HH}^{\rm MHP}=u_{HH}^{\rm HP}=-1$
and rescaled the value of $u_{HP}^{\rm MHP}$ by the factor of $2.3$. This number was 
chosen in Ref.~\cite{tang1} as a result of an analysis for the inter-residue energies of contacts
between hydrophobic amino acids and contacts between hydrophobic and polar 
residues~\cite{mj1} which motivated the relation $2 u_{HP}>u_{PP}+u_{HH}$.
\end{thebibliography}
\end{document}